\documentclass[runningheads]{svmult}
\usepackage{psfig}
\usepackage{makeidx}   % allows index generation
\usepackage{graphicx}  % standard LaTeX graphics tool
\usepackage{subeqnar}  % subnumbers individual equations
\usepackage{physprbb}  % modified textarea for proceedings
\usepackage{multind}
\usepackage{multicol}
\makeindex{s}         
\makeindex{o}

\usepackage{amsmath}
\usepackage{amssymb}

\begin{document}       
\title*{Spiral waves in accretion discs - theory}
\toctitle{Spiral waves in accretion discs - theory}
\titlerunning{Spiral waves in discs}
\author{Henri M.J. Boffin}
\authorrunning{Henri Boffin}
\institute{Royal Observatory of Belgium, Av. circulaire~3, B-1180 Brussels\\
Henri.Boffin@oma.be}
\maketitle
\begin{abstract}
Spirals shocks have been widely studied in the context of galactic dynamics and
protostellar discs. They may however also play an important role in some classes of
close binary stars, and more particularly in \index{s}{cataclysmic variable|(}cataclysmic variables.
In this paper, we review the physics of spirals waves in accretion discs, 
present the results of numerical simulations and consider whether theory can be reconcilied
with observations. 
\end{abstract}

If, in the course of the evolution of a binary system, the separation between
the stars decreases, there comes a point where the gravitational pull of one of
the stars removes matter from its companion. There is mass transfer, through
the so-called \index{s}{Roche equipotential|(}Roche lobe overflow. This is what is believed to happen in
\index{s}{cataclysmic variable}cataclysmic variable stars (CVs; see Warner \cite{Wa95} for a review). In these, a
\index{s}{white dwarf}white dwarf primary removes mass from its low-mass late-type companion which
fills its \index{s}{Roche equipotential}Roche lobe. In this particular case, the decrease in the separation
is due to a loss of \index{s}{angular momentum|(}angular momentum by magnetic braking or by gravitational
radiation.

\section{\index{s}{Roche equipotential}Roche lobe}

Consider a binary system with a primary \index{s}{white dwarf}white dwarf of mass $M_1$, and a
companion of mass $M_2$ with a mean separation $a$. We can define the \index{s}{mass ratio|(}mass ratio,
$q = M_1/M_2$, which, for CVs, is generally smaller than 1.
From Kepler's law, the orbital period, $P_{\rm orb}$, is then:

\begin{equation}
P_{\rm orb}^2 = \frac{4 \pi^2 a^3}{G(M_1 + M_2)} ,
\label{Eq:P}
\end{equation}

$G$ being the gravitational constant, and the masses being expressed in unit of the
solar mass.

\begin{figure}
\begin{center}
\includegraphics[width=.9\textwidth]{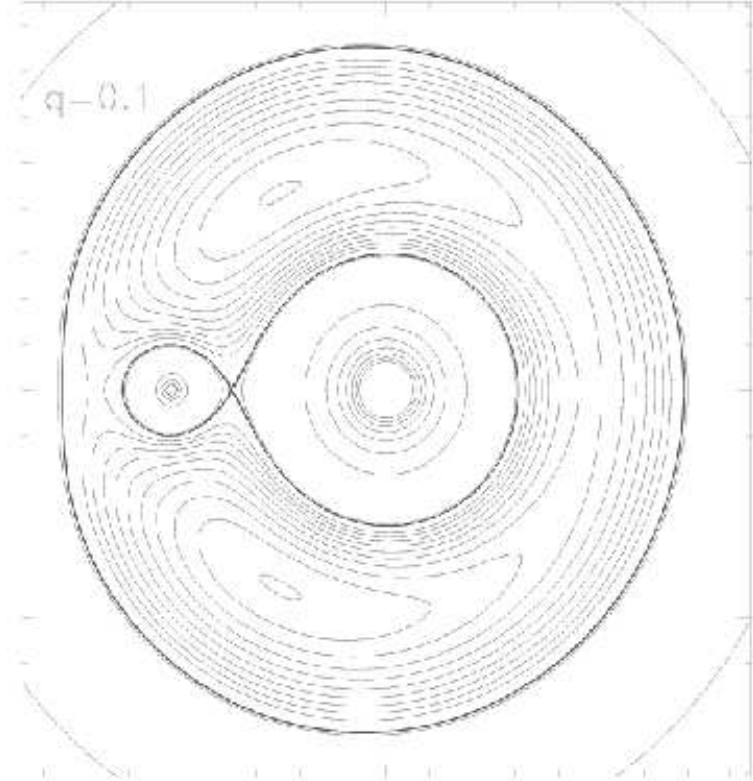}
\end{center}
\vspace{-1cm}
\caption[]{\index{s}{Roche equipotential}Roche equipotentials in a binary system with a \index{s}{mass ratio}mass ratio, $q=0.1$. The Roche
lobes are shown with the heavy lines. The primary is in the middle}
\label{fig:roche}
\end{figure}

In the reference frame rotating with the binary and with the center of mass at the
origin, the gas flow is governed by Euler's equation:

\begin{equation}
\frac{\partial \vec{v}}{\partial t} + (\vec{v \cdot \nabla}) \vec{v} = -  \vec{\nabla} \Phi_r
-2 \vec{\omega} \times \vec{v} - \frac{1}{\rho}  \vec{\nabla} P,
\end{equation}

where $\vec{\omega}$ is the angular velocity of the binary system relative to an
inertial frame, and is normal to the orbital plane with a module
$\omega =  2 \pi / P , ~\rho$ is the density and $P$ is the pressure.
The last term in the right hand side of this equation is thus the gas pressure
gradient, while the second is the \index{s}{Coriolis force}Coriolis force. Here, $\Phi_r$ is the Roche
potential, and includes the effect of both gravitational and \index{s}{centrifugal force}centrifugal forces :

\begin{equation}
\Phi_r = - \frac{GM_1}{|\vec{r}-\vec{r_1}|} - \frac{GM_2}{|\vec{r}-\vec{r_2}|}
- \frac{1}{2} (\vec{\omega} \times \vec{r})^2,
\end{equation}

where $\vec{r_1}, \vec{r_2}$ are the position vectors of the centres of the two stars.
The equipotential surfaces of $\Phi_r$ are shown in Fig. \ref{fig:roche}. It can be
seen that there is a particular equipotential which delimits the two \index{s}{Roche equipotential}Roche lobes of
the stars. The saddle point where the two lobes join is called the inner Lagrange
point, $L_1$. These \index{s}{Roche equipotential}Roche lobes are the key to understanding mass transfer in close
binary systems. Indeed, if one of the two stars fills its \index{s}{Roche equipotential}Roche lobe, then matter
can move into the \index{s}{Roche equipotential}Roche lobe of its companion and be gravitationnally captured by it.
Mass transfer occurs via the so-called \index{s}{Roche equipotential}Roche lobe overflow mechanism.
The size of the \index{s}{Roche equipotential}Roche lobes, $R_{\rm L1}$ and $R_{\rm L2}$,  in unit of the separation is only a
function of the \index{s}{mass ratio}mass ratio, $R_{\rm L2} = a~f(q)$ and $R_{\rm L1} = a~f(q^{-1})$ ,
where an approximate value for $f(q)$ is given by

\begin{equation}
f(q) = \Bigl\{
\begin{array}{cc}
	0.38 + 0.20 \log q &~~~(0.3 \le q < 20), \\
	0.462 \left( \frac{q}{1+q} \right)^{1/3}&~~~(0 < q < 0.3). \\
\end{array}
\label{Eq:rl}
\end{equation}

From Eq.(\ref{Eq:P}) and (\ref{Eq:rl}), one can see that the mean density of a star which fills its \index{s}{Roche equipotential}Roche lobe is a
function of the orbital period only \cite{Wa95}:
\begin{equation}
\bar{\rho} \simeq \frac{107}{P_{\rm orb}^2} ~{\rm gcm^{-3}}, 
\end{equation}
if $P_{\rm orb}$ is expressed in hours. Thus, for the typical orbital periods of CVs,
from 1 to 10 hours, the mean density obtained is typical of lower main-sequence stars.

%%%%%%%%%%%%%%%% FIGURE 1 %%%%%%%%%%%%%%%%%%%%%%%%%%%
%%%%%%%%%%%%%%%%%%%%%%%%%%%%%%%%%%%%%%%%%%%%%%%%%%%%%

%%%%%%%%%%%%%%%% FIGURE 2 %%%%%%%%%%%%%%%%%%%%%%%%%%%
\begin{figure}
\begin{center}
\includegraphics[width=.9\textwidth]{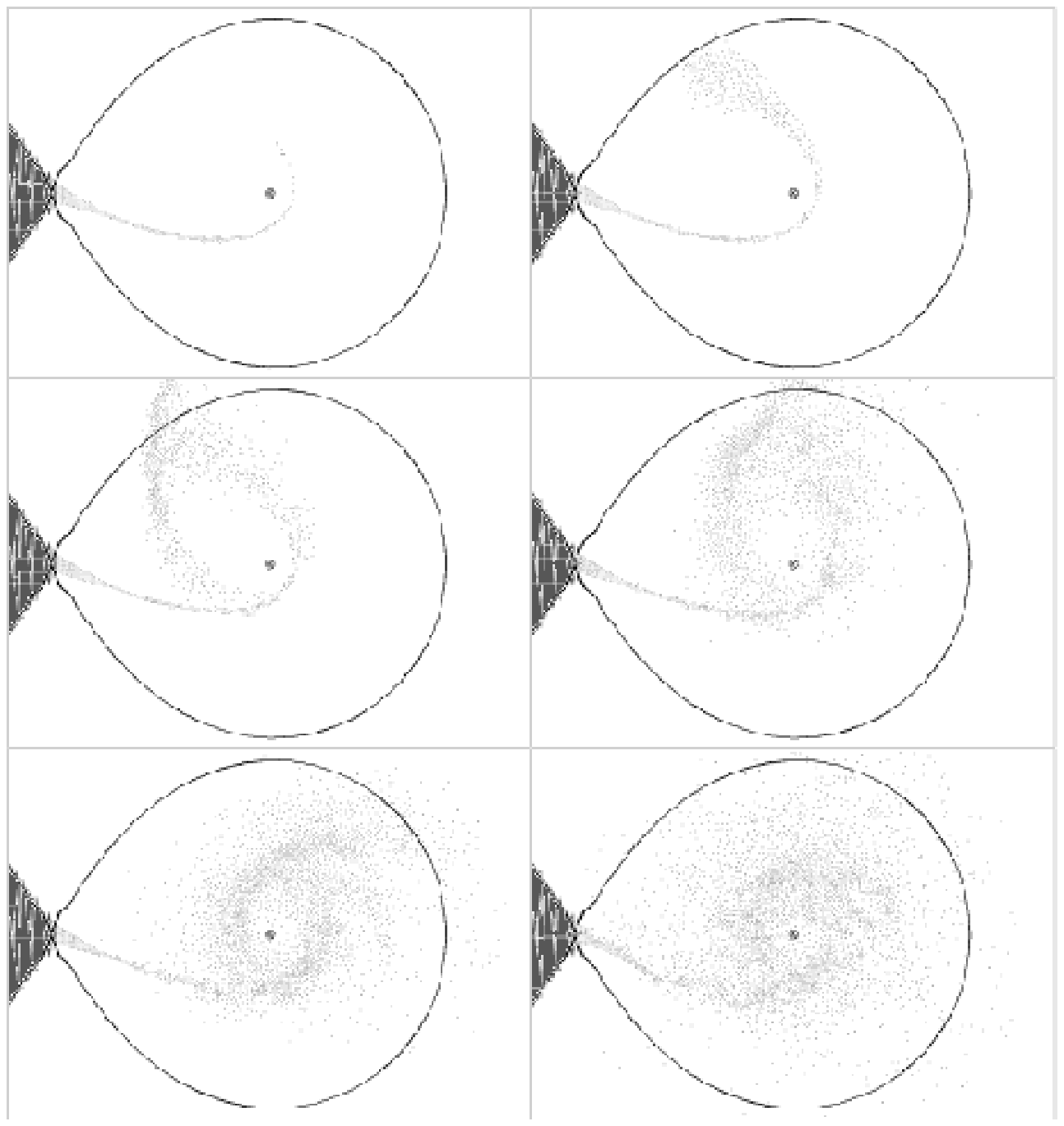}
\end{center}
\vspace{-1cm}
\caption[]{The formation of an accretion disc by \index{s}{Roche equipotential}Roche lobe overflow. The low
mass transferring secondary is at the left. Mass flows through the inner
Lagrange L$_1$ point towards the \index{s}{white dwarf}white dwarf primary}
\label{fig:discform}
\end{figure}
%%%%%%%%%%%%%%%%%%%%%%%%%%%%%%%%%%%%%%%%%%%%%%%%%%%%%

\section{Disc formation}

Matter which is transfered through the $L_1$ point to the companion has a rather
high specific \index{s}{angular momentum}angular momentum with respect to the later, $b^2_1 \omega$.
Here, $b_1$ is the \index{s}{distance}distance of the inner \index{s}{Lagrangian point}Lagrangian point to the centre of the
primary and can be obtained with the following fitted formula :
\begin{equation}
b_1 = ( 0.500 - 0.227 \log q) ~a.
\end{equation}
%Results for various \index{s}{mass ratio}mass ratio are shown in Table \ref{table:fits}.

Once the gas comes close to the primary, it will mainly feel its gravitational force
and follow a \index{s}{Keplerian orbit}Keplerian orbit, for which the circular velociy is given by
\begin{equation}
v_\phi (r) = \sqrt{\frac{GM_1}{r}},
\end{equation}
and the associated \index{s}{angular momentum}angular momentum, $r ~v_\phi (r)$. If we equal this to the angular
momentum of the gas transfered through the \index{s}{Roche equipotential}Roche lobe, we can define the
circularization radius
$R_{\rm circ} = (1+q) (b_1/a)^4 a$. Gas coming from the companion will thus form a
ring of radius approximately $R_{\rm circ}$, provided this is larger than the radius
of the accreting object.
For \index{s}{cataclysmic variable}cataclysmic variable stars, we typically have $P_{\rm orb}= 1-10$ hours,
$M_1 \approx M_\odot$ and $q < 1$.
Therefore, $a \lesssim 3 R_\odot$ and $R_{\rm circ} \approx 0.1-0.3 ~a $.
By comparison, the primary \index{s}{white dwarf}white dwarf has a radius about
0.01 $R_\odot$. Thus, in non-\index{s}{cataclysmic variable!magnetic}magnetic CVs, the mass transfer will always lead to the
formation of such a ring.
This ring will then spread out by viscous processes: because energy is lost,
the gas will move deeper into the gravitational well until it reaches the primary
and accretion will occur (see \cite{fr92}). To conserve \index{s}{angular momentum}angular momentum, the outer part will have
to move further away. An accretion disc forms. The formation of an accretion disc
is pictured in Fig. \ref{fig:discform} as obtained from numerical simulations.

In principle, the disc could expand forever. In a binary system, this is not
possible however because of the torque exerted by the companion. The radius of
the disc is therefore limited by the tidal radius, where the tides 
induced by the \index{s}{secondary star}secondary star truncate the disc. The \index{s}{angular momentum}angular momentum is then
transferred back to the orbital motion. Paczynski \cite{Pacz77} has computed the
maximum size of a disc in a binary system by following the orbits of a set of
particles and deriving the largest non-intersecting orbit.
This orbit would represent the maximum
size of an accretion disc, when neglecting the effect of \index{s}{viscosity}viscosity and pressure.
This is called the disc's tidal truncation radius.
Typically, the \index{s}{accretion disc!radius}disc radius is limited to 0.7-0.9 the \index{s}{Roche equipotential}Roche lobe radius.

\section{Viscosity}

As we have seen, viscous processes are at play to explain to formation of accretion
discs. It is this \index{s}{viscosity}viscosity which will also ensure that matter transfered by the
companion can be accreted onto the primary.
In the thin disc approximation, i.e. when the disc half thickness $H$ is much
smaller than the radius $r$ at each radius, $H/r << 1$, and in a \index{s}{steady state}steady state,
\begin{equation}
\label{Eq:Mdot}
\nu \Sigma = \frac{\dot M}{3 \pi} \left[ 1 -
\left(\frac{R_{\rm wd}}{r}\right)^{1/2} \right],
\end{equation}
where $\nu $ is the effective kinematic \index{s}{viscosity}viscosity, $\Sigma$ the surface density and
$\dot M$ the mass \index{s}{accretion rate}accretion rate (e.g. \cite{fr92}).
The origin of this \index{s}{viscosity}viscosity is as yet unknown. Shakura \& Sunyaev \cite{sha73} used
 a parametric formulation to hide in a parameter, $\alpha$, our lack of knowledge:
\begin{equation}
\label{Eq:alpha}
\nu = \alpha c_{\rm s} H,
\end{equation}
with $c_{\rm s}$ being the sound speed. Note that in the thin disc approximation,
\begin{equation}
\label{Eq:thin}
\frac{H}{r} \sim \frac{c_{\rm s}}{v_\phi} \equiv \frac{1}{\mathcal{M}},
\end{equation}
with $\mathcal{M}$ the Mach number in the disc.
In the framework of the turbulent \index{s}{viscosity}viscosity mechanism, the Shakura \& Sunyaev
prescription can be understood in writing the kinematic \index{s}{viscosity}viscosity as the product of
the turbulent velocity, $v_{\rm t}$ and the
typical eddy size, $l_{\rm ed}$: $ \nu \sim v_{\rm t} l_{\rm ed}$. 
The eddy size cannot be larger than the disc scale height, thus $l_{\rm ed} \lesssim H$.
Moreover, in order to avoid shocks, the turbulent velocity must be subsonic,
$v_{\rm t} \lesssim c_{\rm s}$. Thus, $\alpha$ must be smaller or equal
to 1.
There is no reason however for $\alpha$ to be constant throughout the disc.

By using Eq.~(\ref{Eq:Mdot}) and the direct relation between the mass \index{s}{accretion rate}accretion rate
 and the \index{s}{luminosity}luminosity of the disc, it is possible to have an estimate of the amount of
\index{s}{viscosity}viscosity present in the accretion disc of \index{s}{cataclysmic variable}cataclysmic variables.
In a typical \index{s}{cataclysmic variable!dwarf nova}dwarf nova, the disc is observed to brighten by about five magnitudes
for a period of days every few months or so. According to the thermal instability
model, which is the most widely accepted model to explain these \index{s}{outburst}outbursts, the disc 
flips from a low accretion, cool state in \index{s}{quiescence}quiescence to a high
accretion hot state at \index{s}{outburst}outburst (e.g. \cite{Tout20}).
It is therefore
generally believed that $\alpha \simeq 0.01$ in quiescent \index{s}{cataclysmic variable!dwarf nova}dwarf nova, while $\alpha$
 is typically 0.1-0.3 in \index{s}{cataclysmic variable!dwarf nova}dwarf novae in \index{s}{outburst}outburst or in \index{s}{cataclysmic variable!nova-like}nova-like stars. This is
clearly too large by several order of magnitudes for standard molecular \index{s}{viscosity}viscosity.
More serious candidates are therefore turbulent \index{s}{viscosity}viscosity, \index{s}{magnetic stress}magnetic stresses and the
Balbus-Hawley and Parker magnetic instabilities.

Although the common mechanims invoked for this anomalous \index{s}{viscosity}viscosity are thought
to be local, hence the Shakura \& Sunyaev prescription, it may be possible that
some global mechanism is acting as a sink  for \index{s}{angular momentum}angular momentum. In this case,
one could still use the above prescription if we now use an {\it effective} $\alpha$
parameter. This will be the case for \index{s}{spiral arms|(}spiral shocks which we will discuss in the rest
of this review.

\section{Spiral shocks}

Although it was already known that accretion discs in close binary systems can lose
\index{s}{angular momentum}angular momentum to an orbiting exterior companion through tidal interaction
\cite{li76,PP77}, it is Sawada, Matsuda \& Hachisu \cite{sa86a,sa86b}
who showed, in their 2D inviscid numerical simulations of accretion discs in a
binary of unit \index{s}{mass ratio}mass ratio, that \index{s}{spiral arms}spiral shocks could form which propagate to very
small radii. Spruit \cite{sp87,Sp89} and later Larson \cite{Larson90} made
semi-analytical calculations which were followed by numerous - mostly 2D -
numerical simulations \cite{bo99,go97,Hara99,mak99a,mak99b,Mat87,ma90,mat99a,mat99b,ro89,ro93,sa87,spr87,yu97}. 
An historic overview can be found in Matsuda et al. \cite{Ma2000}.

Savonije, Papaloizou \& Lin \cite{sav94} presented both linear and non-linear
calculations of the tidal interaction of an accretion disc in close binary systems.
The linear theory normally predicts that \index{s}{spiral arms}spiral waves are generated at Lindblad
resonances in the disc. A $m:n$ Lindblad resonance corresponds to the case
where $n$ times the disc angular speed is commensurate to $m$ orbital angular speed:
 $n \Omega = m \omega$,
i.e.
%$n^2 M_1 / r^3 = m^2 (M_1 + M_2) / a^3
$ r = \left( \frac{1}{1+q} (\frac{n}{m})^2 \right)^{1/3} a $.
In fact, in their study of tidal torques on accretion discs in binary systems
with extreme \index{s}{mass ratio}mass ratios, Lin \& Papaloizou \cite{LinPa79} already observed the
spiral pattern. In their case, this pattern was indeed due to the 2:1
Lindblad resonance which can fall inside the \index{s}{Roche equipotential|)}Roche lobe and inside the disc if
the \index{s}{mass ratio}mass ratio is small enough.  However, the typical  \index{s}{mass ratio}mass ratios and \index{s}{accretion disc!radius}disc radius of
\index{s}{cataclysmic variable}cataclysmic variable stars does not allow the centre of such resonances
to be in the disc. 
For the 2:1 resonance to lay inside the disc requires $q <
0.025$, a value too small for most \index{s}{cataclysmic variable}cataclysmic variables altough possible for
some low-mass \index{s}{X-ray binary}X-ray binaries. The 3:1 resonance can be located inside the disc
for $q < 0.33$, hence for most of \index{s}{cataclysmic variable!SU UMa class}SU UMa stars (see e.g. \cite{WK91}).

In their study, however, Lin \& Papaloizou \cite{LinPa79} showed that the resonant
effect is significant over a region
\begin{equation}
\Delta x \sim \left( \frac{\nu ~ r}{v_\phi ~ r_s^2} \right) ^{1/3} ~r,
\end{equation}
where $r_s$ is the position of the resonance. Using $\nu = \alpha c_s H$, this
leads to $\Delta x \propto c_s^{2/3}$.
Thus, as also found by Savonije et al. \cite{sav94}, even in CVs with
larger \index{s}{mass ratio}mass ratios, the centre of the 2:1 resonance can still be thought of as
lying in the vicinity of the boundaries of the disc and because the resonance
has a finite width that increases with the magnitude of the sound speed, it can
still generate a substantial wave-like spiral response in the disc, but only if
the disc is large, inviscid and the Mach number is smaller than about 10.  For
larger Mach number, more typical of \index{s}{cataclysmic variable}cataclysmic variables, however, Savonije et
al. consider that wave excitation and propagation becomes ineffective and
unable to reach small radii at significant amplitude.

We will look more closely to this later on. As for now, we will follow Spruit \cite{sp87}
to show why \index{s}{spiral arms}spiral waves can be thought of as an effective \index{s}{viscosity}viscosity.
Spiral waves in discs have been sudied extensively in the context of galactic
dynamics and of protostellar discs. Such waves carry a negative agular
momentum. Their dissipation leads to accretion of the fluid supporting the waves
onto the central object.

%%%%%%%%%%%%%%%% FIGURE 3 %%%%%%%%%%%%%%%%%%%%%%%%%%%
\begin{figure}
\begin{center}
\includegraphics[width=.45\textwidth]{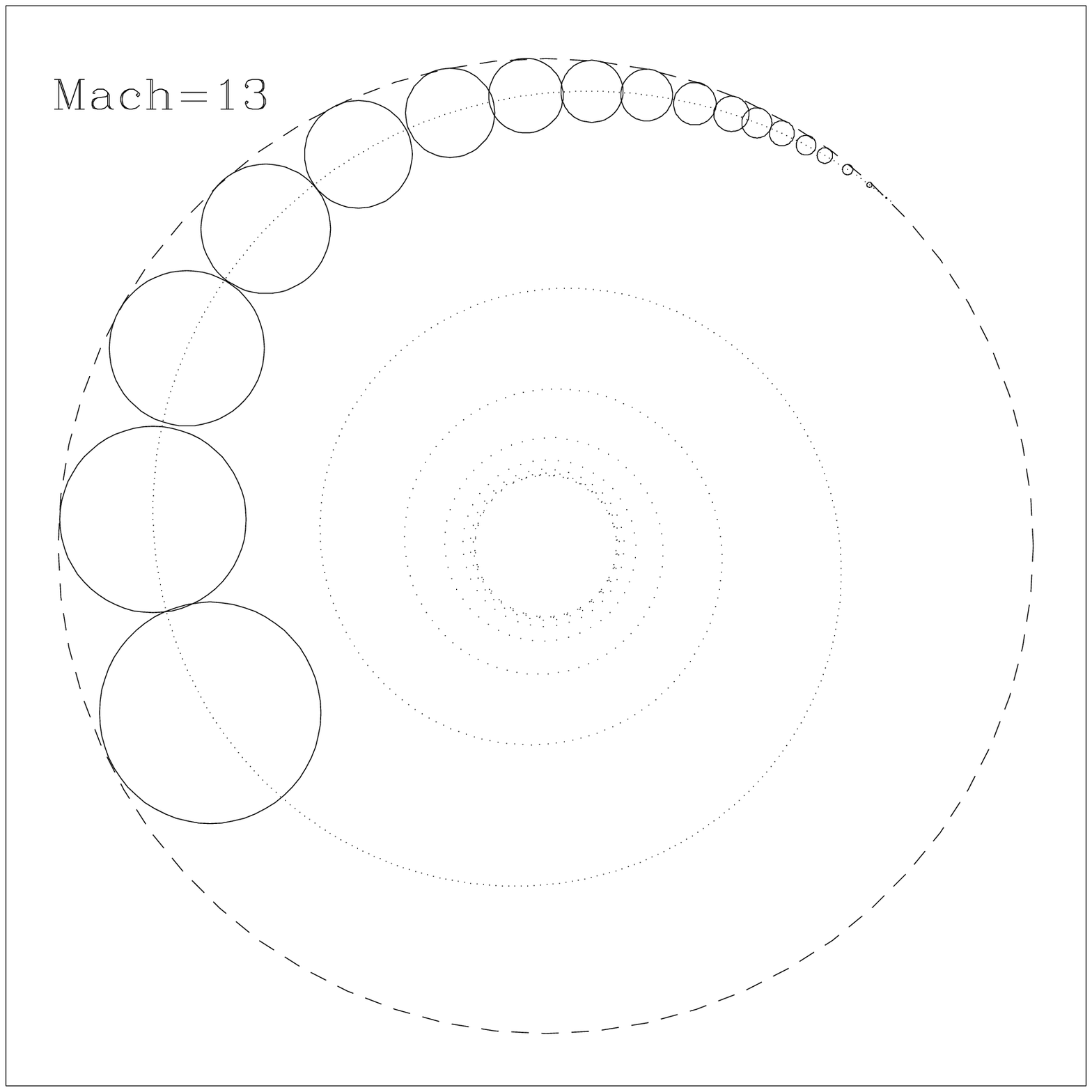}
\includegraphics[width=.45\textwidth]{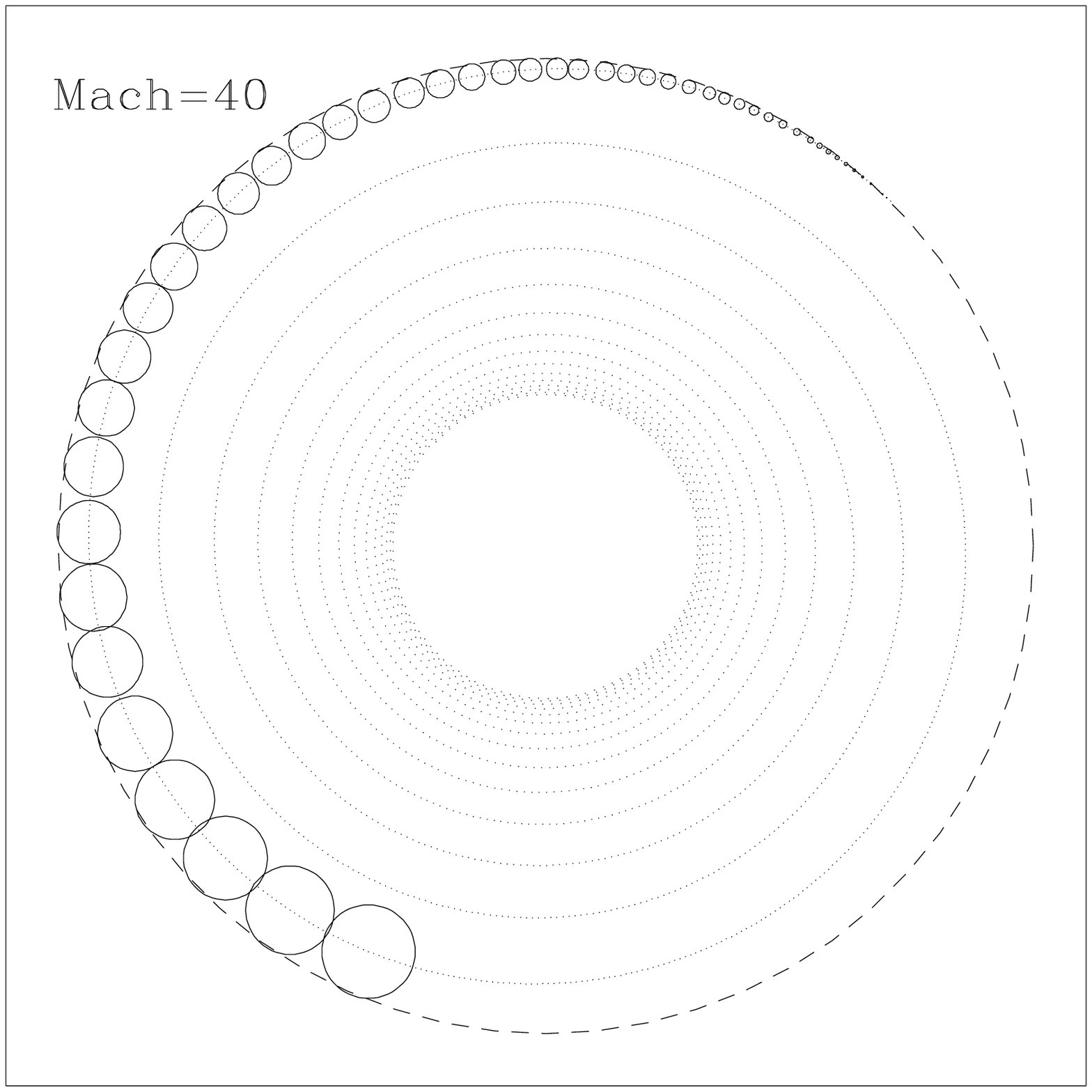}
\end{center}
\vspace{-7mm}
\caption[]{Illustration of the fact that the angle of the spiral pattern depends
on the Mach number. In the hot disc (left), the perturbance propagates faster and the
spiral is more open than in the cold disc (right)
}
\label{fig:spiral}
\end{figure}
%%%%%%%%%%%%%%%%%%%%%%%%%%%%%%%%%%%%%%%%%%%%%%%%%%%%%
Consider some disturbance at the outer disc edge. As the generated wave 
propagates inward, it is being wound up by the differential \index{s}{Keplerian orbit}Keplerian
rotation in the disc into a trailing spiral pattern (Fig.~\ref{fig:spiral}). 
The wave frequency $\sigma$ for an azimuthal
wavenumber $m$ in the comoving frame is
$\sigma = \omega - m \Omega \simeq  -m \Omega$, because we can neglect the 
much lower orbital frequency, $\omega$.
The conserved wave action is given by
\begin{equation}
S_{\rm w} = \frac{1}{2} \int \frac{\rho v^2_{\rm w}}{\sigma} dV
\end{equation}
where $v_{\rm w}$ is the amplitude of the wave and the integration is carried out
over the volume of the wave packet. The \index{s}{angular momentum}angular momentum of the wave,
given by $j = m S_{\rm w}$, 
is conserved, giving:
\begin{equation}
j \simeq - \frac{1}{2} \int \frac{\rho v^2_{\rm w}}{\Omega} dV.
\end{equation}
Hence, it is negative. This can be understood because the tidally excited spiral
pattern rotates with the binary angular speed which is smaller than the angular
speed of the gas in the disc.

%%%%%%%%%%%%%%%% FIGURE 4 %%%%%%%%%%%%%%%%%%%%%%%%%%%
\begin{figure}[h]
\begin{center}
\includegraphics[width=.45\textwidth]{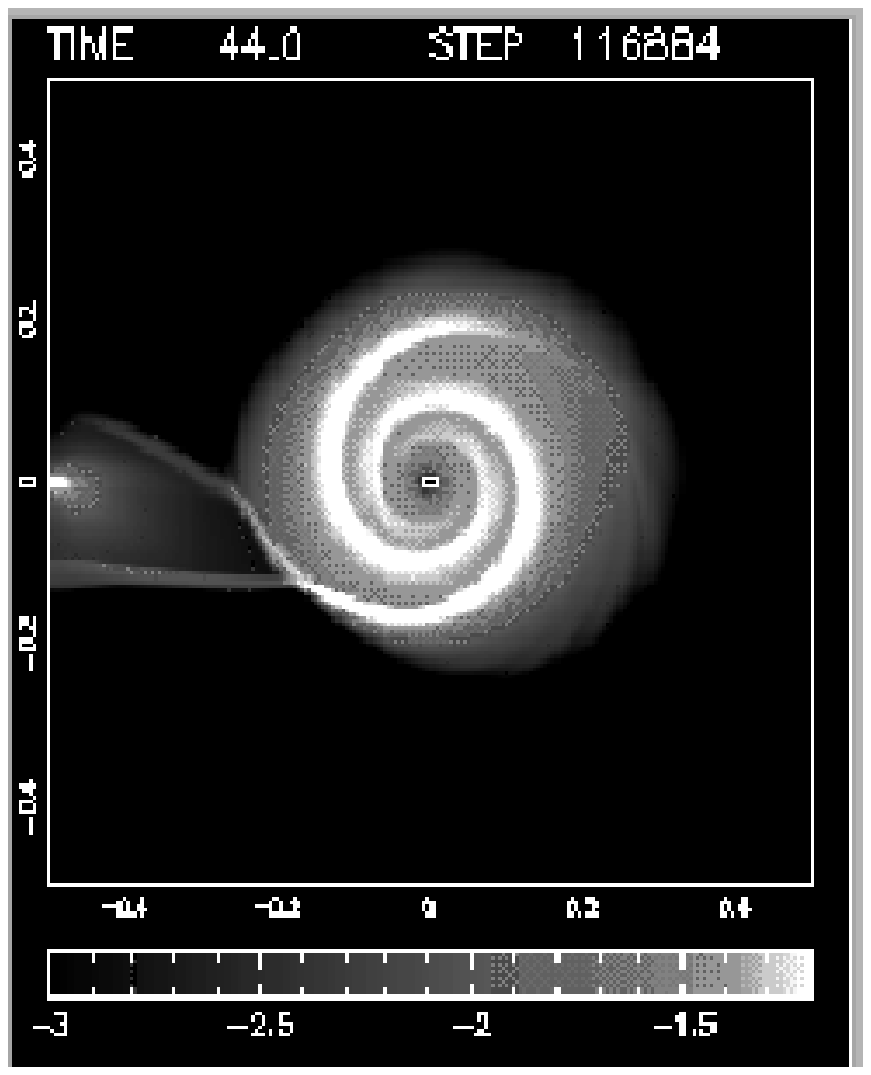}
\includegraphics[width=.45\textwidth]{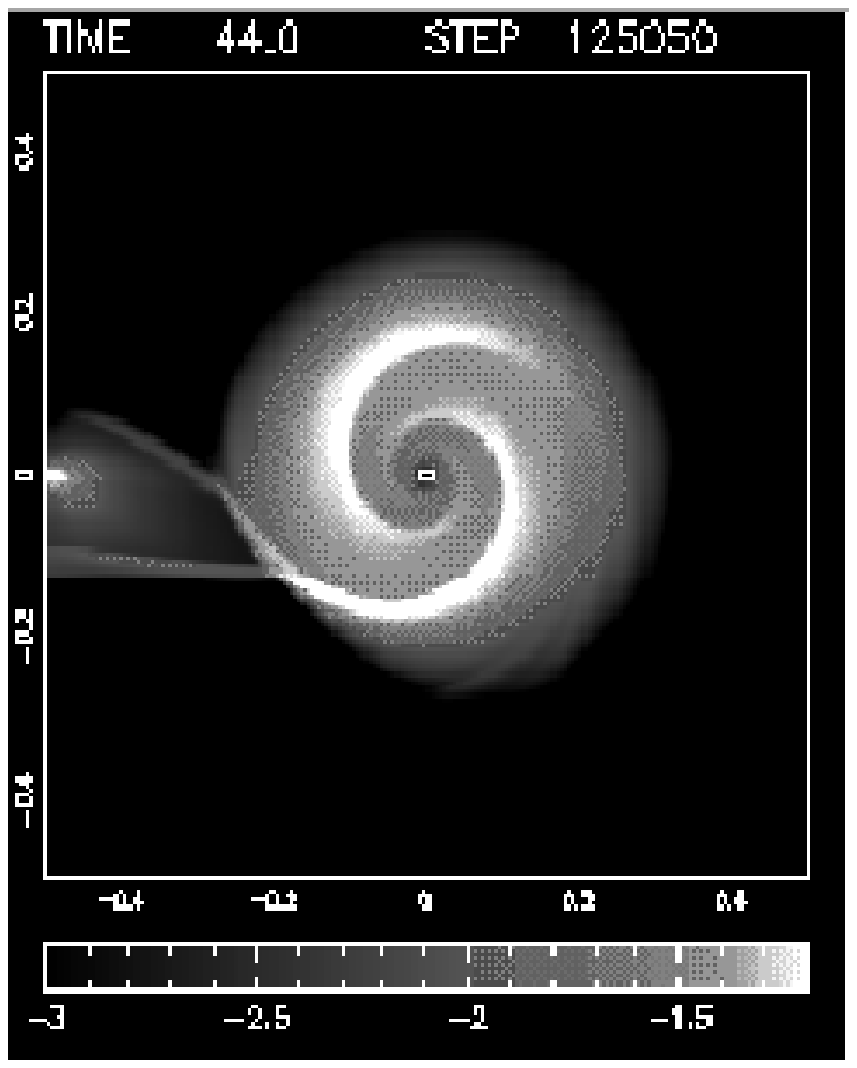}\\
\includegraphics[width=.45\textwidth]{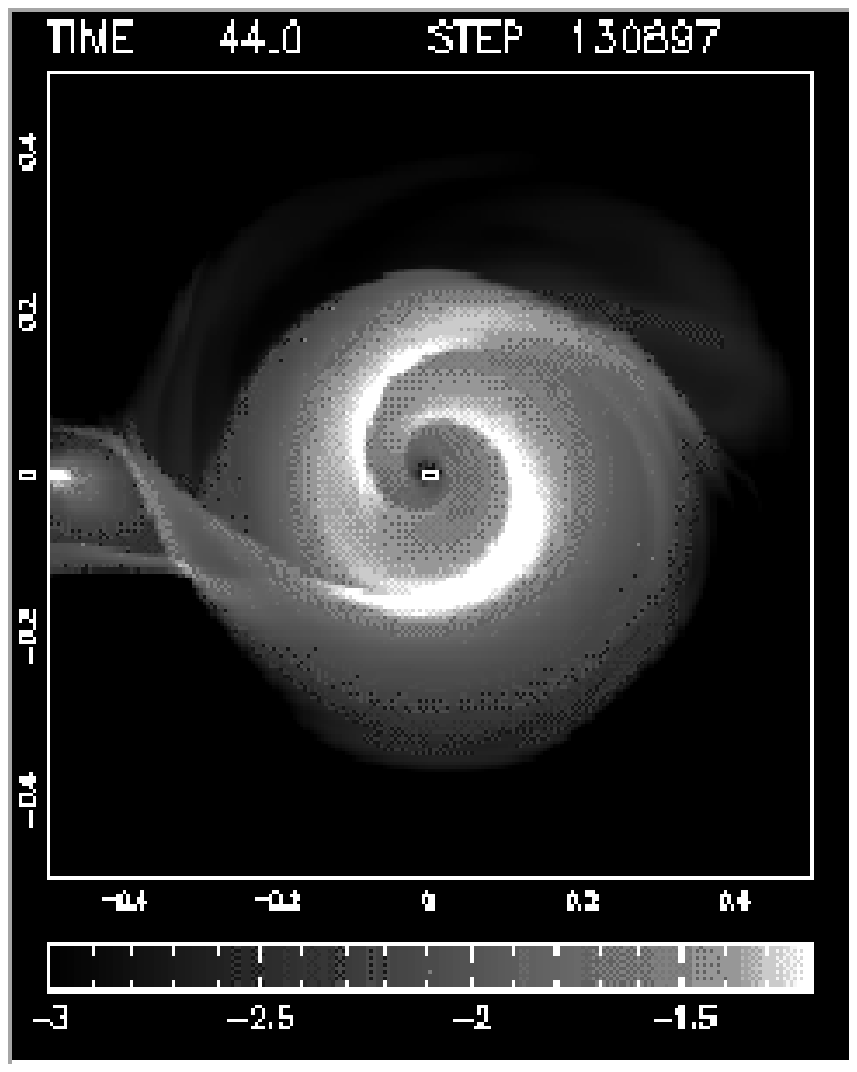}
\includegraphics[width=.45\textwidth]{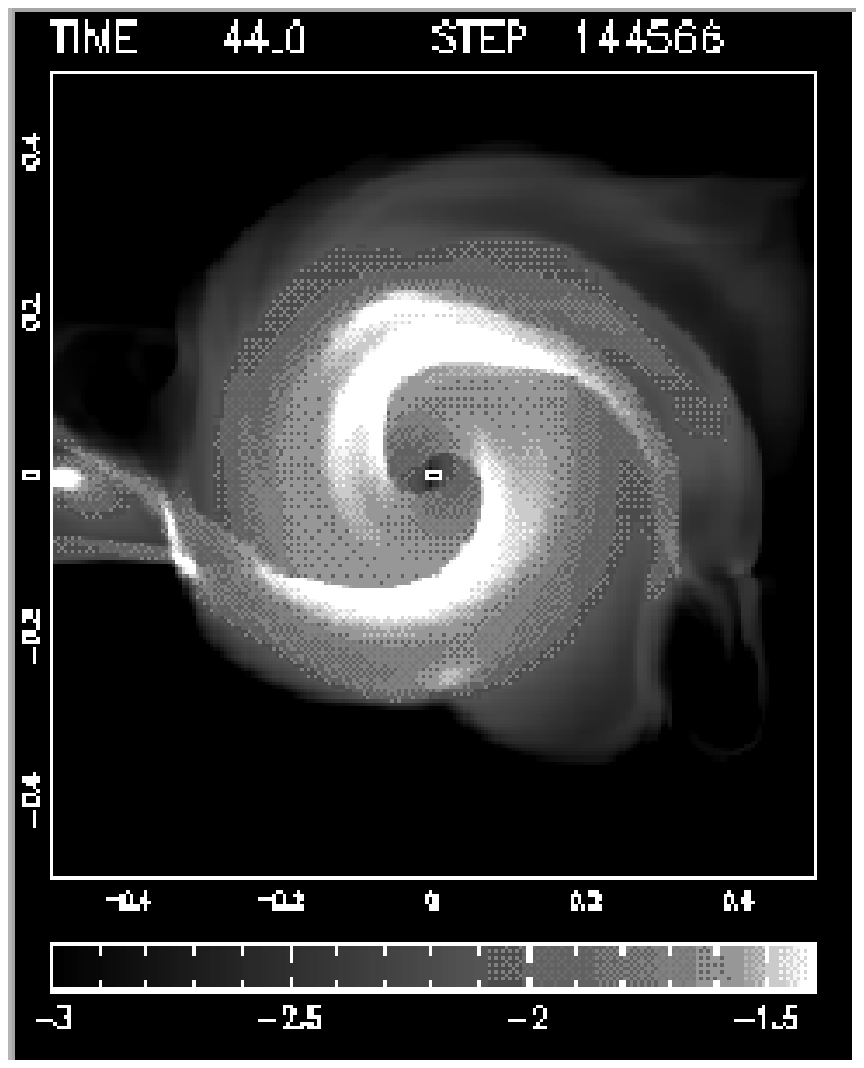}
\end{center}
\caption[]{Density plots of 2D finite-difference simulations (in the case of a
\index{s}{mass ratio}mass ratio of 1)
for different value of $\gamma$ = 1.01 (upper left), 1.05, 1.1, 1.2 (lower right). The
scale is logarithmic and the results are shown at about 7 orbital periods \cite{mak98}
}
\label{fig:fd2d}
\end{figure}
%%%%%%%%%%%%%%%%%%%%%%%%%%%%%%%%%%%%%%%%%%%%%%%%%%%%%
Because of the \index{s}{differential rotation}differential rotation, the amplitude of the wave increases as it 
propagates inwards. 
Indeed, the radial extent of the wave packet, $\Delta R$, is 
proportional to the sound speed, 
$\Delta R \propto c_{\rm s}$, while  
the volume of the wave packet is $ V = 2 H \times 2 \pi R \Delta R $. 
Now, because $j$ is conserved, Spruit obtains
\begin{equation}
v_{\rm w} \propto \left( \frac{\Omega}{\rho r H c_{\rm s}} \right) ^{1/2} \propto r^{-11/16},
\end{equation}
where in the last approximation, we made use of the relations valid for a thin disc. 
The wave therefore steepens into a shock. 
Dissipation in the shock lead to a loss of \index{s}{angular momentum|)}angular momentum in the disc, hence to accretion.

The opening angle of the spirals is related directly to the temperature of the disc as, 
when the shock is only of moderate strength, it roughly propagates at sound speed. 
This is shown schematically in Fig.~\ref{fig:spiral} for two values of the Mach number, 
where we have assumed that the perturbation propagates radially at a speed of 1.3 times the
Mach number \cite{blo99,La89,sp87}. 
Thus the angle between the shock surface and the direction of the orbital motion is 
of the order $\tan ~\theta = c_s / v_\phi = 1 / \mathcal{M}$. 
In the approximation of an adiabiatic equation of state, often used in numerical simulations, 
where the temperature soon reaches a value given by 
$T=0.5 (\gamma - 1) T_{\rm vir}$, with $T_{\rm vir}$ being the virial temperature, 
the angle becomes $\sqrt{0.5 \gamma (\gamma - 1.)}$. 
Thus, low $\gamma$ discs will have more tightly wound spirals than large $\gamma$ ones. 
This is indeed was is shown by two-dimensional simulations (\cite{mak98,Ma2000}; see also Fig.~\ref{fig:fd2d}).

\section{Observational facts}

The most prevalent and successful model to explain \index{s}{cataclysmic variable!dwarf nova}dwarf nova \index{s}{outburst|(}outbursts is
the disc instability model based on a \index{s}{viscosity}viscosity switch related to the \index{s}{ionisation}ionisation
of hydrogen in the disc. It is an hysteresis cycle, in which the disc switches
back and forth between a hot \index{s}{optically thick}optically thick high \index{s}{viscosity}viscosity state - the \index{s}{outburst}outburst - 
and a cool \index{s}{optically thin}optically thin low \index{s}{viscosity}viscosity state - \index{s}{quiescence}quiescence.
The \index{s}{accretion disc!radius}disc radius increases at \index{s}{outburst}outburst and then after maximum, decreases exponentially.
In \index{o}{U Gem}U Gem for example, a clear increase in the radius is seen at \index{s}{outburst}outburst :
the radius of the disc is of the order of $0.4 a$ at maximum and then it decreases on a
timescale of tens of days to $0.28 a$ \cite{Wa95}.
While for \index{o}{EX Dra}EX Dra, Baptista \& Catalan \cite{BC99} found  the \index{s}{accretion disc!radius}disc radius to be 
$0.30 a$ in \index{s}{quiescence}quiescence and $0.49 a$ in \index{s}{outburst}outburst.

Global disc evolution models reproduce the main properties of observed
\index{s}{outburst}outbursts, but only if the efficiency of transport and dissipation is less in
\index{s}{quiescence}quiescence than during \index{s}{outburst}outburst \cite{Ca93}.
This seems in agreement with \index{s}{spiral arms}spiral shocks. 
Indeed, the tidal force is a very steep function of $r/a$, hence the spirals rapidly
become weak at smaller disc sizes. 
Spiral shocks are produced in the outer regions of the disc by the tides
raised by the \index{s}{secondary star}secondary star. During the \index{s}{outburst}outburst, the disc expands, and
its outer parts feel the gravitational attraction of the \index{s}{secondary star}secondary star more
efficiently, leading to the formation of \index{s}{spiral arms}spiral arms.

Having these general ideas in mind, we can now summarise the results from observations
(see the review by Danny Steeghs in this volume):
\begin{itemize}
\item The spiral pattern has been established in several \index{s}{cataclysmic variable}cataclysmic variables
for a wide range of emisson lines
\item The \index{s}{spiral arms}spiral arms appear right at the start of the \index{s}{outburst}outburst and persist during
\index{s}{outburst}outburst maximum for at least 8 days, i.e. several tens of orbital periods
\item The structure is fixed in the corotating frame of the binary and corresponds
to the location of tidally driven \index{s}{spiral arms}spiral waves
\item During \index{s}{quiescence}quiescence, the spiral pattern is no longer there but the disc remains
asymmetric
\end{itemize}
These results can be exactly interpreted in the \index{s}{spiral arms}spiral shock theory.
They do not prove necessarily, however, that \index{s}{spiral arms}spiral shocks are the main \index{s}{viscosity}viscosity 
mechanism in the disc. There is indeed a problem similar to the egg and the hen: who 
begon? In order to have well developed spirals, one needs a large disc, which is 
due to an increase in \index{s}{viscosity}viscosity !

%%%%%%%%%%%%%%%% FIGURE 5 %%%%%%%%%%%%%%%%%%%%%%%%%%%
\begin{figure}
\begin{center}
\includegraphics[width=.9\textwidth]{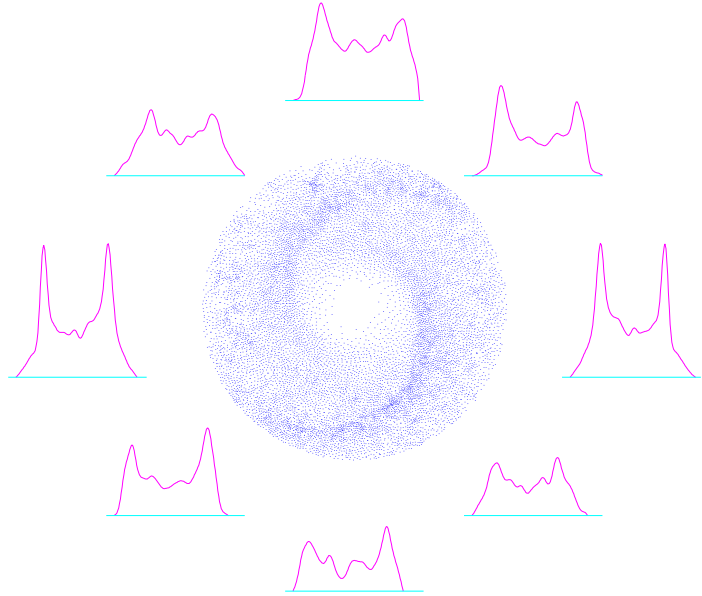}
\end{center}
\vspace{-2cm}
\caption[]{ An accretion disc showing \index{s}{spiral arms}spiral arms as obtained by \index{s}{smoothed particle hydrodynamics}SPH
simulations.
When viewed from different angles ({\it i.e.} at different orbital phases), the
emission-line profile is different.
When these lines are recorded at several orbital phases, one can then construct
a spectrogram which can be inverted, using a maximum \index{s}{entropy}entropy method, to give a
doppler \index{s}{Doppler map}tomogram
\label{fig:spectra}
}
\end{figure}
%%%%%%%%%%%%%%%%%%%%%%%%%%%%%%%%%%%%%%%%%%%%%%%%%%%%%

%%%%%%%%%%%%%%%% FIGURE 6 %%%%%%%%%%%%%%%%%%%%%%%%%%%
\begin{figure}
\begin{center}
\includegraphics[width=.99\textwidth]{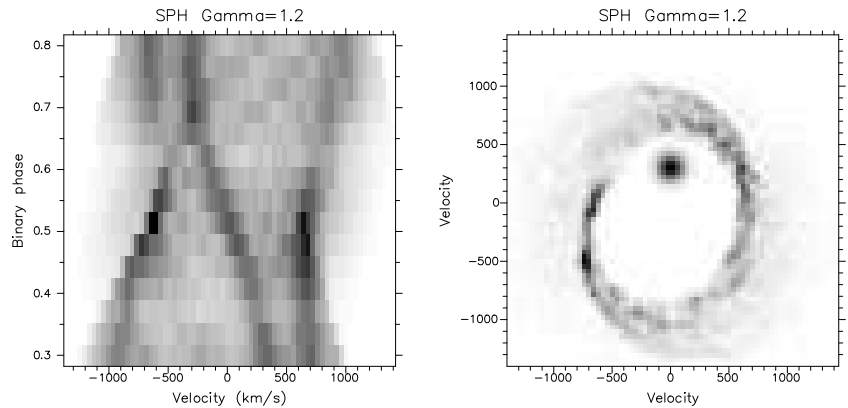}
\end{center}
\vspace{-5mm}
\caption[]{Binary phase-velocity map (spectrogram) and \index{s}{Doppler map}Doppler map
corresponding to
the simulation of Fig.~5
}
\label{fig:tomo}
\end{figure}
%%%%%%%%%%%%%%%%%%%%%%%%%%%%%%%%%%%%%%%%%%%%%%%%%%%%%

%%%%%%%%%%%%%%%% FIGURE 7 %%%%%%%%%%%%%%%%%%%%%%%%%%%
\begin{figure}
\begin{center}
\includegraphics[width=.99\textwidth]{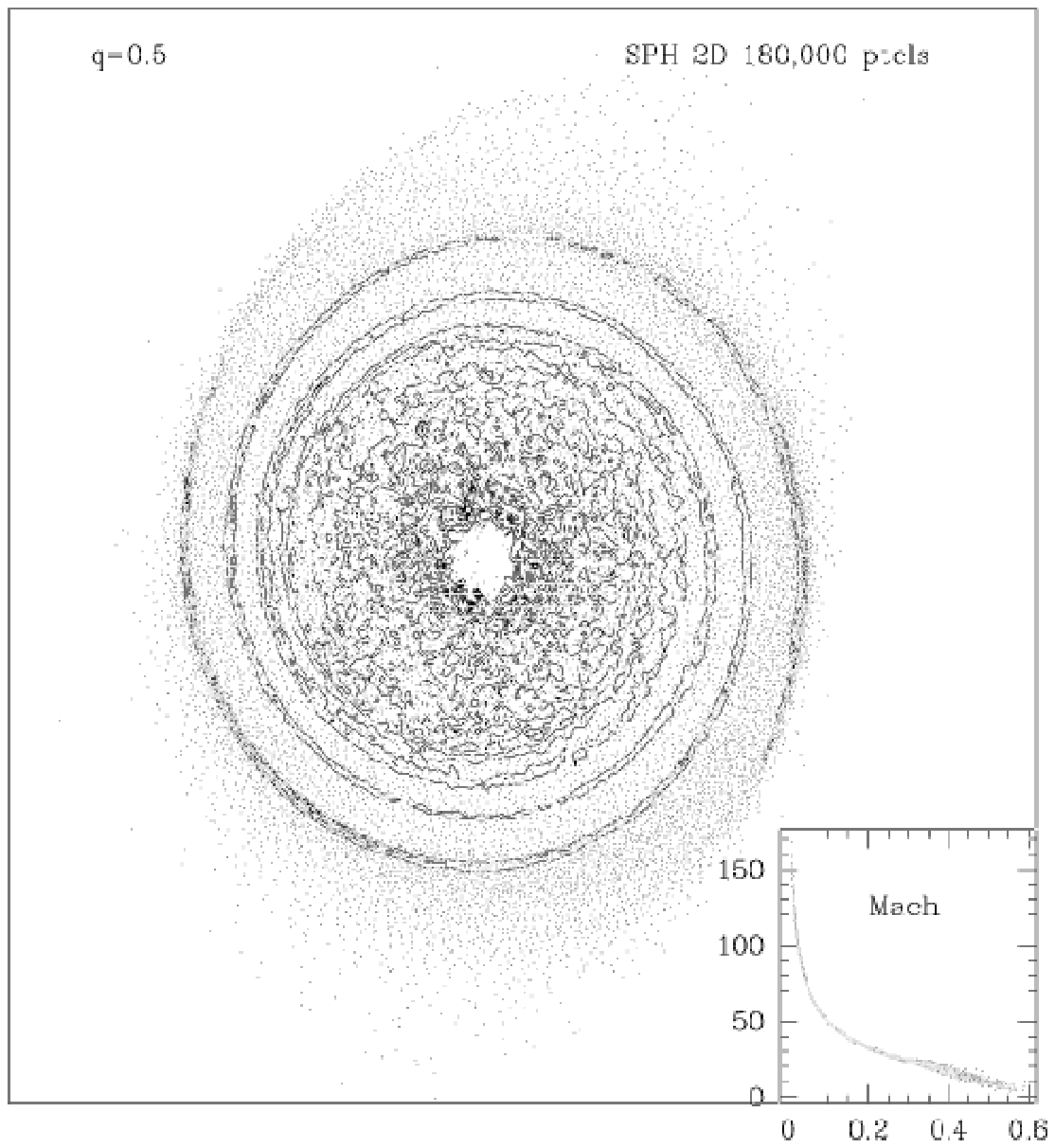}
\end{center}
\vspace{-1cm}
\caption[]{Result of a high-resolution 2D \index{s}{smoothed particle hydrodynamics}SPH simulation 
of an isothermal ($c_s = 0.05$) accretion disc in a binary system with 
a \index{s}{mass ratio}mass ratio of 0.5. The density contours are plotted 
over the particles positions. The inset shows the variation
of the Mach number with the \index{s}{distance}distance from the \index{s}{white dwarf}white dwarf.
}
\label{fig:sph2discool}
\end{figure}
%%%%%%%%%%%%%%%%%%%%%%%%%%%%%%%%%%%%%%%%%%%%%%%%%%%%%

%%%%%%%%%%%%%%%% FIGURE 8 %%%%%%%%%%%%%%%%%%%%%%%%%%%
\begin{figure}[bpt]
\begin{center}
\includegraphics[width=.4\textwidth]{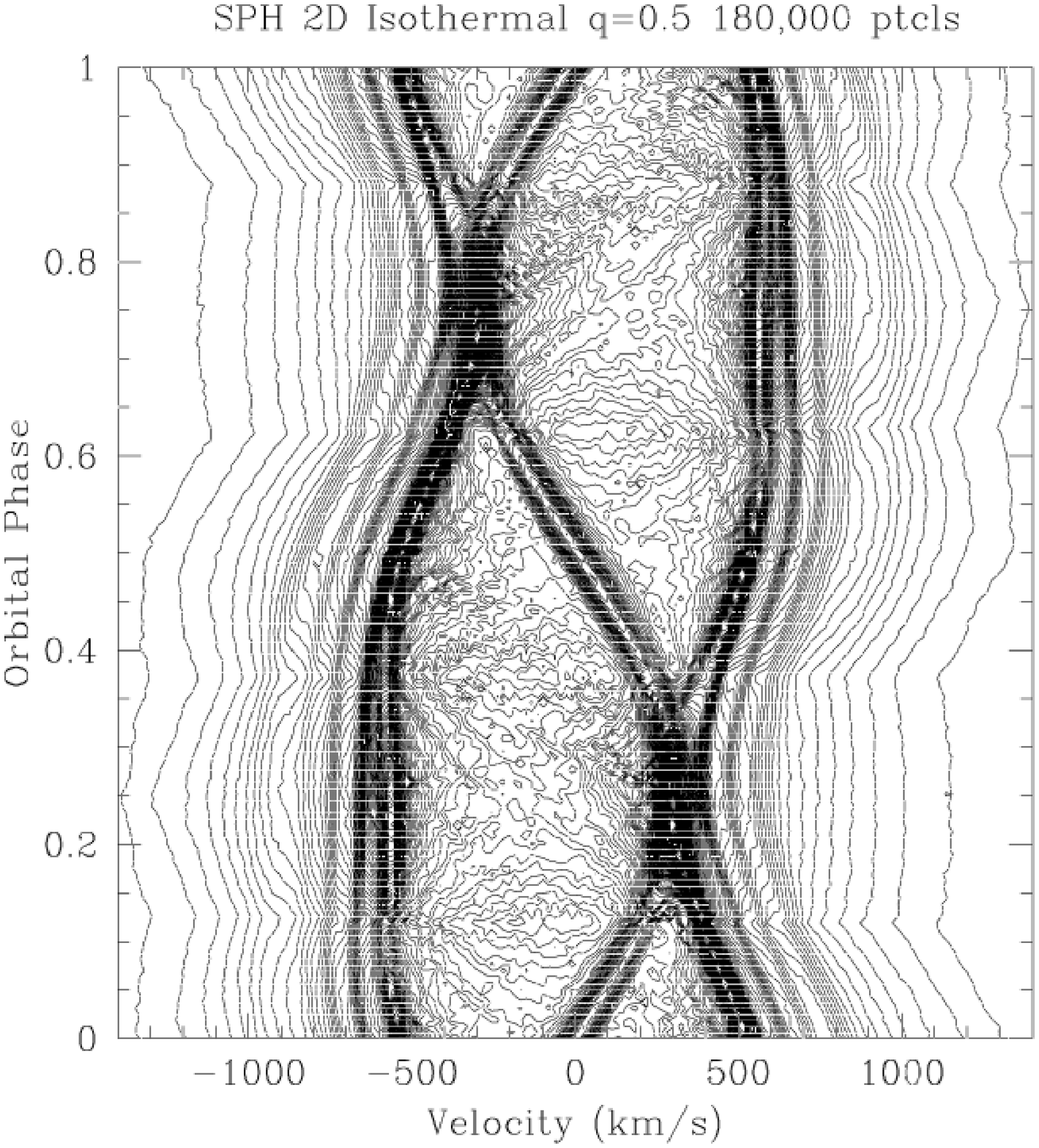}
\includegraphics[width=.4\textwidth]{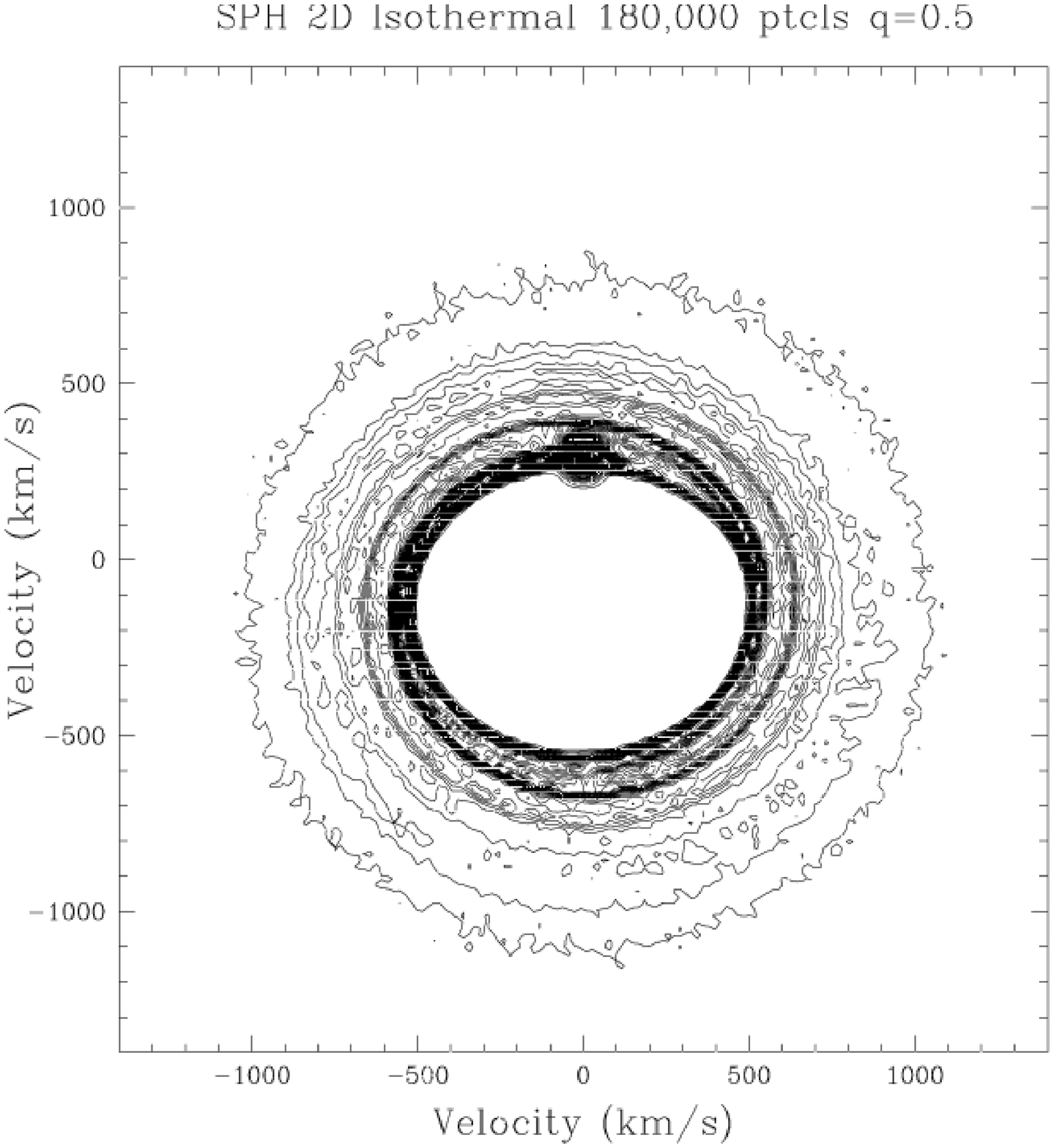}
\end{center}
\caption{Reconstructed trailed spectrogram and \index{s}{Doppler map}Doppler map
from the numerical simulation shown in Fig.~\ref{fig:sph2discool}
}
\label{fig:discooltomo}
\end{figure}
%%%%%%%%%%%%%%%%%%%%%%%%%%%%%%%%%%%%%%%%%%%%%%%%%%%%%

%%%%%%%%%%%%%%%% FIGURE 9 %%%%%%%%%%%%%%%%%%%%%%%%%%%
\begin{figure}
\begin{center}
\includegraphics[width=.9\textwidth]{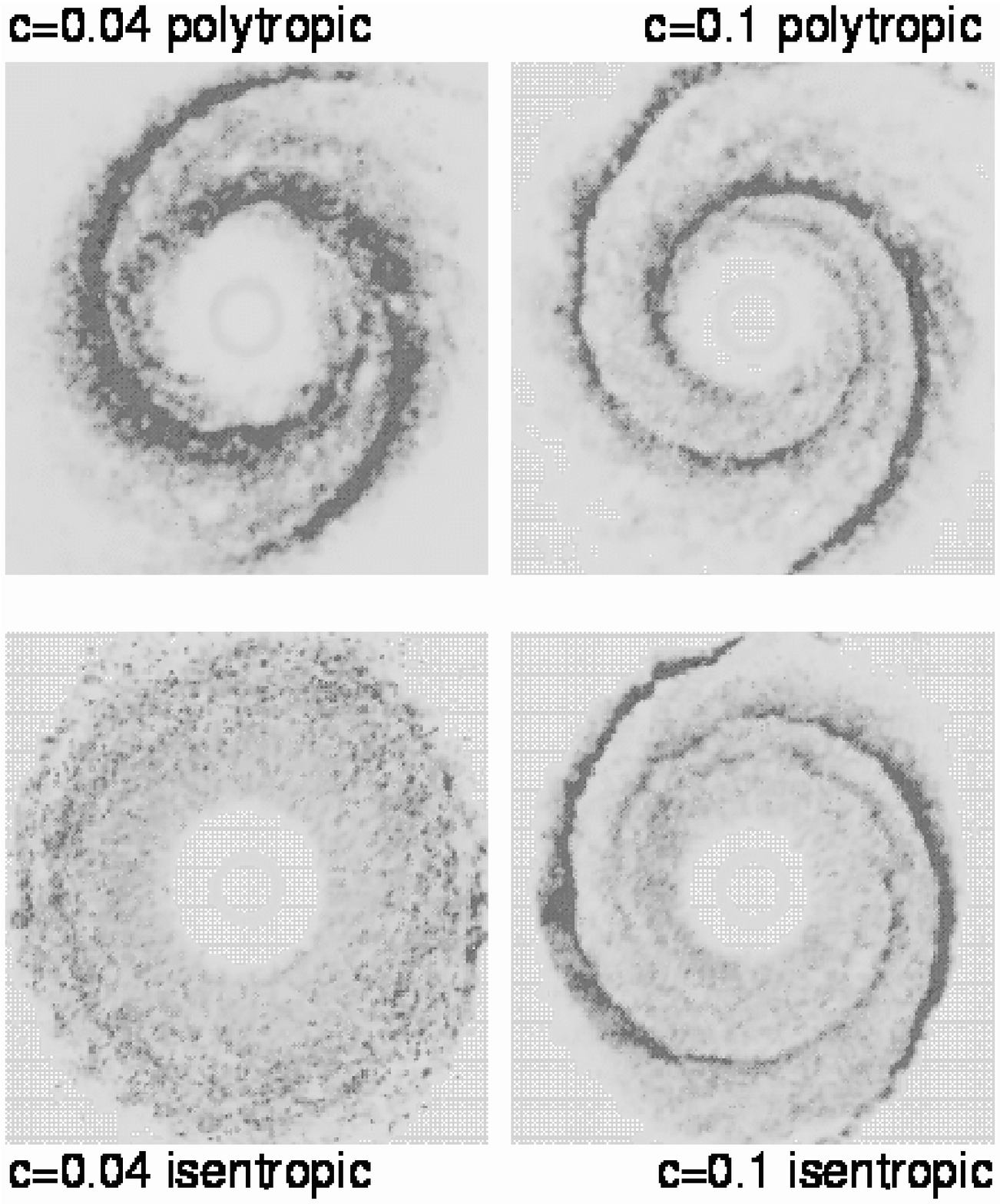}
\end{center}
\vspace{-5mm}
\caption{Comparison between our \index{s}{smoothed particle hydrodynamics}SPH results for the isentropic and
polytropic equation of state for two values of c$_o$=0.1 and c$_o$=0.04\label{isen}}
\label{fig:isen}
\end{figure}
%%%%%%%%%%%%%%%%%%%%%%%%%%%%%%%%%%%%%%%%%%%%%%%%%%%%%

\section{Numerical simulations}

As can already be seen from Fig.~\ref{fig:spiral} and \ref{fig:fd2d}, in order to explain 
the widely open spirals seen in the \index{s}{Doppler map}Doppler maps of \index{o}{IP Peg}IP Peg \cite{SHH97}, one require 
a rather hot disc. 
Figure~\ref{fig:spectra} shows for example the results of a 2D numerical simulation, using
the Smoothed Particle Hydrodynamics (\index{s}{smoothed particle hydrodynamics}SPH) method, of an accretion disc in a binary
system with a \index{s}{mass ratio}mass ratio of 0.5, as observed for \index{o}{IP Peg}IP Peg. Here, a polytropic equation of state
(EOS) has been used with a polytropic index, $\gamma=1.2$.
In Fig.~\ref{fig:spectra}, we show the different emission lines profiles corresponding to 
the various orbital phases for such a disc. Those profiles can then be presented in 
a trailed spectrogram as seen in Fig.~\ref{fig:tomo}. With numerical simulations, we have
the advantage of having all the dynamical information of the flow and we can therefore
easily construct a \index{s}{Doppler map}Doppler map which can then be compared to those observed. The Doppler
map corresponding to this $\gamma=1.2$, 2D simulation is also shown in  Fig.~\ref{fig:tomo}. 
Note that we have artificially added a spot at the location of the secondary to allow 
more direct comparisons with real observations. 
It is obvious that even though this simulation was not the result of any tuning, the 
qualitative agreement between this calculated \index{s}{Doppler map}Doppler map and spectrogram, and the one
first observed by Steeghs et al. \cite{SHH97} for \index{o}{IP Peg}IP Peg is very good.
This however calls for some discussion. Indeed, in simulations using a polytropic EOS, 
as is the case here, the disc evolve towards the virial temperature and are therefore
unphysically hot. There is thus an apparent contradiction between numerical simulations
and observations. This was indeed the conclusions of the first few attempts 
to model the observed spirals in \index{o}{IP Peg}IP Peg in \index{s}{outburst}outburst.
For comparison purposes, we present in Fig.~\ref{fig:sph2discool} and \ref{fig:discooltomo}, 
the results of a very high resolution isothermal simulation. 

Godon, Livio \& Lubow \cite{go98} presented two-dimensional disc simulations using
Fourier-Chebyshev spectral methods and found that the spiral pattern resembles
the observations only for very high temperatures. 
This was already the conclusion of Savonije et al. \cite{sav94} who claimed that 
spirals could not appear in the colder disc of \index{s}{cataclysmic variable}cataclysmic variables.

We have therefore run another set of simulations, with our \index{s}{smoothed particle hydrodynamics}SPH code, where we use an 
``isentropic'' equation of state, instead of a polytropic one.
By ``isentropic'', we mean a barotropic EOS, $P = K \rho ^\gamma$, and keeping $K$ constant.
The heating due to the viscous processes is supposed to be instantaneously radiated away.
In this case, the temperature remains always very close to the initial value.
The results of isentropic runs are compared with those of polytropic runs using
$\gamma$=1.2 and two values for the initial sound speed: 0.1 and 0.04 times the orbital 
velocity (Fig.~\ref{fig:isen}).
While, in the two polytropic runs, \index{s}{spiral arms}spiral structures are clearly present with a very 
similar pitch angle, there is a clear distinction between the two isentropic runs: 
the 0.1 case produces well defined
\index{s}{spiral arms}spiral arms which are more tighly wound than in the equivalent polytropic case, 
while in the 0.04 case,
\index{s}{spiral arms}spiral structure can hardly be seen at all.
The difference clearly lies in the Mach number of the flow: in the polytropic cases, 
the  Mach number is well below 10 for both a sound speed of 0.04 and 0.1,
while in the isentropic cases, the Mach number goes from 20 at the outer edge of the disc to more
than 100 inside when the sound speed is 0.04, but goes from 10 to only 30  when the sound speed is 0.1.
Therefore, the observations seem to require a hot disc with a wide spiral.
But, are \index{s}{cataclysmic variable!dwarf nova}dwarf novae discs hot or cold ?
During \index{s}{quiescence}quiescence, the disc will have a temperature of the order of 10,000 K.
This corresponds to a sound speed of the order of 15 km/s, or in our units 0.03.
Thus, one should compare observations with our isentropic case with sound speed 0.04, and we therefore predict that no spiral should be seen.
In \index{s}{outburst}outburst however, a \index{s}{cataclysmic variable!dwarf nova}dwarf nova will have a \index{s}{temperature profile}temperature profile corresponding to that of 
the steady-state solution for a viscous disc (e.g. \cite{Wa95}), and the temperature will 
be closer to $10^5$ K, {\it i.e.} a sound speed of roughly 45 km/s (in our units, 0.1).
In this case, the Mach number will be around 20 and one is allowed to compare observations with the isentropic case with a sound speed of 0.1.
Thus, one should clearly expect to see \index{s}{spiral arms}spiral structures in \index{s}{cataclysmic variable!dwarf nova}dwarf novae in \index{s}{outburst}outburst but not in quiesence.
%The observations of IP Peg by Steeghs et al. , Harlaftis \& Steeghs (1998) and Marsh \& Horne (1990) show that this is indeed the case.
%We therefore conclude that, contrary to some claims in the literature, there is no contradiction between numerical simulations of \index{s}{spiral arms}spiral structures in accretion discs and observations.
%This is also confirmed by the recent study of Steeghs and Stehle (1999).

Similarly, Stehle (1998) performed thin disc calculations where the full set of time dependent
hydrodynamic equations is solved on a cylindrical grid. The \index{s}{accretion disc!thickness}disc thickness is
explicitly followed by two additional equations in a one-zone model, allowing
the disc to be vertically in non-equilibrium. The spatial and temporal evolution
of the \index{s}{accretion disc!temperature}disc temperature follows tidal and viscous heating, the latter in the
$\alpha$-ansatz of Shakura \& Sunayev, as well as radiation from the disc surfaces.
The surface temperature is connected to the disc mid-plane temperature using
Kramers opacities for the vertical radiation transport. In this sense, it is a
much more elaborated model than the classical isothermal or adiabatic approximation
for the equation of state.
Steeghs \& Stehle (1999) use the grid of disc calculations of Stehle (1998) to
construct Doppler \index{s}{Doppler map}tomograms for a binary with \index{s}{mass ratio}mass ratio $q=0.3$ and orbital period
$P=2.3$ hours. They considered two models, one more typical of quiescent CVs,
with $M \simeq 15-30$ and an $\alpha$-type shear \index{s}{viscosity}viscosity of 0.01, and another,
representative of \index{s}{outburst}outbursting discs, with $M \simeq 5-20$ and $\alpha=0.3$.
Effective temperature ranged from less then $10^4$ K in the outer part of the cold disc,
to values between $2\,10^4$ and $5\,10^4$ K for the hotter disc. The cold disc
was rather small (due to the low \index{s}{viscosity}viscosity), varying in radius between
0.55 and 0.65 $r_{\rm L1}$, while the hot disc is pushed by the increased \index{s}{viscosity}viscosity
to larger radii of 0.6-0.8 $r_{\rm L1}$.
For the high Mach number accretion discs, they find that the \index{s}{spiral arms}spiral shocks are so
tightly wound that they leave few fingerprints in the emission lines, the double
peaks separation varying by at most 8\%. For the accretion disc in \index{s}{outburst}outburst, however,
they conclude that the lines are dominated by the emission from an $m=2$ spiral
pattern in the disc, resulting in converging emission line peaks with a cross over
near phases 0.25 and 0.75.
It has to be noted that in the simulations of Stehle, it is the presence of a large
initial shear \index{s}{viscosity}viscosity (in the form of an $\alpha$-type parametrization) which
provide the viscous transport required to setup a large hot disc in which a strong
two armed spiral patter forms. This remark will take all its relevance in view of
the discussion in Sect.~\ref{sect:qui}. 

For example, Godon et al. \cite{go98} used a value of $\alpha = 0.1$,
so that viscous spreading was less efficient compared to the simulations of
Stehle. As a consequence, their discs were smaller by up to 50 \% compared to the
hot disc model of Steeghs \& Stehle, hence the \index{s}{tidal effect}tidal effect was less severe.

\section{Angular momentum transport}

As noted above, the parameter $\alpha$ introduced by Shakura and Sunayev \cite{sha73}
refers to some local unknown \index{s}{viscosity}viscosity. In the case of \index{s}{spiral arms}spiral shocks, which is 
a global phenomenon, we
can still refer to an {\it effective} $\alpha$ that would give the same mass
\index{s}{accretion rate}accretion rate in an $\alpha$-disk model that one finds in numerical
simulations.

The standard
$\alpha$-disk theory gives the mass \index{s}{accretion rate}accretion rate in terms of $\alpha$
as $\dot M = 3\pi\alpha c_s H\Sigma$ [see Eq.~(\ref{Eq:Mdot}) and (\ref{Eq:alpha})].
Using Eq.(\ref{Eq:thin}), Blondin \cite{blo99}
finds an equation for $\alpha$:
\begin{equation}
\alpha_{\rm eff} = {2\over 3}{v_\phi \langle v_r\rangle\over c_s^2},
\end{equation}
where $\langle v_r \rangle$ is a density-weighted average of the \index{s}{radial velocity}radial velocity:
\begin{displaymath}
\langle v_r\rangle = {\int_0^{2\pi}\Sigma v_r d\phi \over
         \int_0^{2\pi}\Sigma  d\phi }.
\end{displaymath}

It has to be noted that the accretion time scale can be estimated (e.g. \cite{Sp20}) :
$$ \tau _{\alpha} \sim \frac{1}{\alpha_{\rm eff} \Omega} \left(\frac{v_\phi}{c_s}\right)^2 ~.$$
Thus this time scale can only be followed for rather hot discs with high value of 
$\alpha_{\rm eff}$ and therefore, for cold discs, the evolution can usually be 
followed until the wave pattern is staionary but not long enough for the accretion 
process to reach a \index{s}{steady state}steady state. The wave pattern on the other hand is stable in 
only a few sound crossing time.

From his study
 of self-similar models of very cool discs, Spruit \cite{sp87} obtains
$\alpha _{\rm eff} = 0.013 \left( \frac{H}{r} \right) ^{3/2} $. As in cataclysmic
variables, $\frac{H}{r} \leq 0.1$, this leads to very low values
($10^{-4} - 10^{-3}$), which was the reason why several people dismissed spiral
shocks as a viable efficient accretion mechanism. Spruit himself, however, in
his paper, insists that this must not be the final word: "For most common \index{s}{mass ratio}mass ratios 
$0.1 < q < 1$, the forcing by the companion is so strong however that the resulting 
\index{s}{spiral arms}spiral shock therefore have a strength much above the self similar value over a large 
part of the disc". In their inviscid - but adiabatic - simulations, 
Matsuda et al. (1987) obtained values of the effective $\alpha$ up to 0.1, hence 
large enough to explain mass accretion observed in \index{s}{outburst}outbursting \index{s}{cataclysmic variable|)}cataclysmic variables. 
In fact, Spruit \cite{sp87} and Larson \cite{La89} obtained a relation between 
the effective $\alpha$ and the radial Mach number $M_r$ at disc mid-plane. 
Larson \cite{La89}, for example, obtains
$ \alpha _{\rm eff} \simeq 0.07 (M_r^2 - 1)^3$ for isothermal discs. 
Note that for an isothermal disc, $M_r^2$ is equal to the compression ratio. 
With $M_r$
being of the order of 1.3 to 1.5, this typically implies values of $\alpha _{\rm eff} \simeq 0.02 - 0.14$.
Numerical simulations by Blondin \cite{blo99} seem to confirm this, as in his isothermal
simulations, values as high as 0.1 are obtained near the outer edge of the disc.
For very cold discs, he even found values close to 1.
The accretion efficiency however decreases very sharply and reaches value
below $10^{-3}$ in the part of the disc closer than $0.1 a$. For a hotter disc,
if the value of $ \alpha _{\rm eff}$ is about 0.1 in the outer part of the disc,
it stays above 0.01 well inside the disc.
Larson \cite{La89} goes on to predict that the strength of tightly
wound \index{s}{spiral arms}spiral shocks in a cold, thin disk should be proportional to
$(\cos\theta )^{3/2}$, where $\theta$ is the trailing angle of the spiral
wave.  Estimating $\cos\theta \approx 1.5c_s/v_\phi$,
Larson \cite{La89} finds an effective $\alpha$ of
\begin{equation}
\alpha \approx 0.026 \left( {c_s\over v_\phi}\right)^{3/2}\, .
\end{equation}
The prediction for an isothermal disk with $c_s=0.25$ is $\alpha
\approx 3.2\times 10^{-3}\, r^{3/4}$, orders of magnitude below that
found in the numerical simulations (e.g. \cite{blo99,Mat87}).
But in these simulations, the spirals are not tightly wound and the 
formula of Larson may not be applicable.

Blondin \cite{blo99}, for example, finds that the maximum
value of $\alpha$, found near the outer edge of the disk, remains
roughly independent of sound speed, while the radial dependence of
$\alpha (r)$ steepens with decreasing sound speed.  In fact, because
the radial decay of $\alpha$ is so steep, he found steady mass
accretion only for radii above $r\approx 0.1a$ in the coldest disk with
an outer Mach number of $\sim 32$.
A further result of his simulations is the relative independence of
\index{s}{spiral arms}spiral waves on the \index{s}{mass ratio}mass ratio in the binary system.  The strength of
the two-armed \index{s}{spiral arms}spiral shocks at their origin near the outer edge of
the disk was fairly constant in all of his models.  Blondin believes this has the
consequence that, despite all else, one can be confident that the
effective $\alpha$ in the outer regions of an accretion disk in
a binary system is (at least) $\sim 0.1$.

%%%%%%%%%%%%%%%%%%%%%%%%%%

\section{Spirals in quiescence ?}
\label{sect:qui}

By using results from shearing-box simulations as an input for a global
numerical model designed to study disc instabilities, Menou \cite{Me20} shows
that as the disc goes into \index{s}{quiescence}quiescence, it suffers a runaway cooling. Hence, in
\index{s}{quiescence}quiescence, the disappearance of self-sustained \index{s}{magnetohydrodynamic}MHD turbulence is
guaranteed. As accretion is known to occur during \index{s}{quiescence}quiescence in cataclysmic
variables, another transport mechanism  must operate in the discs. The
rapid disc expansion observed during the \index{s}{outburst}outbursts of several \index{s}{cataclysmic variable!dwarf nova}dwarf novae is
consistent with \index{s}{magnetohydrodynamic}MHD-driven accretion because it shows that disc internal
stresses dominate transport during this phase \cite{Me20}. On the other hand,
the same discs are observed to shrink between consecutive \index{s}{outburst}outbursts, which is
a signature that transport is dominated by the tidal torque due to the
companion star, at least in the outer regions of the disc during \index{s}{quiescence}quiescence.
When looking at a sample of 6 well studied \index{s}{cataclysmic variable!SU UMa class}SU UMa stars, Menou \cite{Me20}
finds an anti-correlation of the recurrence times with \index{s}{mass ratio|)}mass ratios. This, he
interprets, is evidence that tidal torques dominate the transport in the
quiescent discs. Indeed, the recurrence times of \index{s}{cataclysmic variable!dwarf nova}dwarf novae represent the
time-scales for mass and \index{s}{angular momentum}angular momentum redistribution in the quiescent
discs. For his small sample, the correlation is significant for normal and
super-\index{s}{outburst}outbursts. No correlation is however found for U Gem type \index{s}{cataclysmic variable!dwarf nova}dwarf novae
but this could be because the correlation is masked by other effects.
Menou \cite{Me20} proposes thus the concept of accretion driven by \index{s}{magnetohydrodynamic}MHD
turbulence during \index{s}{outburst}outburst and by tidal perturbations during \index{s}{quiescence}quiescence.

This is, at first sight, a rather astonishing result. Indeed, we have seen that the spirals
will be more developed and more effective when the disc is hotter and
larger, i.e. in \index{s}{outburst}outburst. If the main source of \index{s}{viscosity}viscosity during \index{s}{quiescence}quiescence 
are the \index{s}{spiral arms}spiral shocks, then one might expect that their contribution during
\index{s}{outburst|)}outburst cannot be negligible. But as stated before, one needs first to make
the disc large for the \index{s}{tidal effect}tidal effect to become more important. 
The answer may lie - like always - in a combination
of processes. Spiral arms are indeed very good at transporting angular 
momentum in the outer part of the disc. But they do not succeed generally to 
penetrate deep into the disc. This is were \index{s}{magnetohydrodynamic}MHD turbulence may provide the 
main source of \index{s}{viscosity}viscosity. This scenario clearly needs to be further developed
and tested.

%%%%%%%%%%%%%%%%%%%%%%%%%%

\section{Spiral shocks in other stellar objects}

Even if \index{s}{spiral arms}spiral shocks are not the main driving \index{s}{viscosity}viscosity mechanism in accretion
discs, Murray et al. (1998) argue that they could have another observational
consequence in intermediate \index{s}{cataclysmic variable!polar}polars. Indeed, the presence of \index{s}{spiral arms}spiral waves break
the axisymmetry of the inner disc and tells the accreting star the orbital phase
of its companion. This could put an additional variation in the \index{s}{accretion rate}accretion rate
onto the \index{s}{white dwarf}white dwarf, a variation dependent on the orbital period. This could
explain the observed periodic emission in intermediate \index{s}{cataclysmic variable!polar}polars.

Savonije, Papaloizou \& Lin \cite{sav94} note that although tidally induced
density waves may not be the dominant carrier of mass and \index{s}{angular momentum}angular momentum
throughout discs in CVs, they are more likely to play an important role in
protostellar discs around T Tauri binary stars, where the
Mach number is relatively small. In \index{o}{GW Ori}GW Ori for example, both circumstellar and
circumbinary discs have been inferred from infrared excesses \cite{MatAL91} and
the disc response to the tidal disturbance might be significant.
In the same line of ideas, Boffin et al.
\cite{BofWBFW98,WatBBFW98a,WatBBFW98b,WitBN98} found large tidal waves in
large protostellar discs being induced by the tidal interaction of a passing
star. In this case, the perturbance might be large enough in these
self-gravitating discs to lead to the collapse of some of the gas, thereby
forming new stars as well as brown dwarfs and jovian planets.

Another class of objects where \index{s}{spiral arms|)}spiral arms may play a role is \index{s}{X-ray binary}X-ray binaries
(see the review by E. Harlaftis in this volume). There also, because the discs
are rather hot, the Mach number is much smaller than in CVs, hence their effect
might be more pronounced. An observational confirmation of this would be most
welcome. The work of Soria et al.~\cite{SWH20} is maybe such a first
step. These authors studied optical spectra of the soft X-ray transient
\index{o}{GRO J1655-40}GRO J1655-40. They claim that, during the \index{s}{high state}high state, the Balmer
emission appears to come only from a double-armed region on the disc, possibly
the locations of tidal density waves or spirals shocks.

%\eject
%%%%%%%%%%%%%%%%%%%%%%%%%%%%%%%%%%%%%

\end{document}